\documentclass[twocolumn,aps,pre,superscriptaddress,amsmath,amssymb]{revtex4-2}

\usepackage{siunitx}
\usepackage{gensymb,amssymb}
\usepackage[bottom]{footmisc}
\usepackage{lipsum}
\usepackage[utf8]{inputenc}
\usepackage{graphicx}
\usepackage{dcolumn}
\usepackage{bm}
\usepackage{flushend}
\usepackage{xcolor}
\usepackage{lipsum}
\usepackage{tcolorbox}
\usepackage[shortlabels]{enumitem}

\usepackage{hyperref}
\hypersetup{
    breaklinks=true, 
 colorlinks=true, 
 citecolor=magenta,
  urlcolor= blue, 
  linkcolor= blue, 
  bookmarksopen=true, 
}

\newcommand{\eff}{_{\mathrm{eff}}}
\newcommand{\kth}{k_\theta}
\newcommand{\ehh}{\eta(\theta,t)}
\newcommand{\ett}{\eta(t)}
\newcommand{\wt}{\widetilde{W}}

\begin{document}

\title{Evidence of experimental three-wave resonant interactions \\ between two dispersion branches}

\author{Filip Novkoski}\email[E-mail: ]{filip.novkoski@u-paris.fr}
\affiliation{Universit\'e Paris Cité, CNRS, MSC, UMR 7057, F-75013 Paris, France}
\author{Chi-Tuong Pham}\email[E-mail: ]{chi-tuong.pham@upsaclay.fr}
\affiliation{Universit\'e Paris-Saclay, CNRS, LISN, UMR 9015, F-91405 Orsay, France}
\author{Eric Falcon}\email[E-mail: ]{eric.falcon@u-paris.fr}
\affiliation{Universit\'e Paris Cité, CNRS, MSC, UMR 7057, F-75013 Paris, France}

\begin{abstract}
  We report the observation of nonlinear three-wave resonant interactions
  between two different branches of the dispersion relation of hydrodynamic
  waves, namely the gravity-capillary and sloshing modes. These atypical
  interactions are investigated within a torus of fluid for which the sloshing
  mode can be easily excited. A triadic resonance instability is then observed
  due to this three-wave two-branch interaction mechanism. An exponential growth
  of the instability and phase locking are evidenced. The efficiency of this
  interaction is found to be maximal when the gravity-capillary phase velocity
  matches the group velocity of the sloshing mode. For a stronger forcing,
  additional waves are generated by a cascade of three-wave interactions
  populating the wave spectrum. Such a three-wave two-branch interaction
  mechanism is probably not restricted to hydrodynamics and could be of interest
  in other systems involving several propagation modes.
\end{abstract}

\maketitle

\section{Introduction}
Nonlinear wave interactions occur in a variety of systems, where waves of different wavenumbers and frequencies can exchange energy through nonlinear couplings. Such interactions also form the basis for wave-turbulent regimes, where a whole ensemble of waves with different wavenumbers interact among each other and can exhibit a cascade of energy from large to small scales~\cite{Zakharov1992}.

The case of three-wave interactions is prevalent in many areas, such as plasma physics~\cite{Stenflo1994,Hoijer1970}, nonlinear optics~\cite{Armstrong1962}, Rossby
waves~\cite{Connaughton2010}, and even mechanical systems such as suspended cables~\cite{Guo2016} or thin rings~\cite{Kovriguine1998}. For hydrodynamic surface waves, three-wave interactions have been extensively studied in the case of gravity-capillary waves~\cite{Haudin2016,Cazaubiel2019} and hydroelastic waves~\cite{Deike2017}. In the case of gravity waves, four-wave interactions dominate~\cite{Bonnefoy2016,Liu2022}. However, considering one-dimensional (1D) deep-water propagation leads at the leading order to five-wave resonant interactions for either pure capillary waves~\cite{Ricard2021} or pure gravity
waves~\cite{Dyachenko1995,Lvov1997}.

Three-wave systems can also be the source of instability~\cite{Hasselmann1967}. Depending on the
nonlinear coupling between the three waves, a single ``mother'' wave can give rise to two
``daughter'' waves, that then grow exponentially in amplitude. The waves of this triadic resonant
instability (TRI) satisfy both resonances in wavenumber and frequency and has been widely observed
in internal waves in stratified flows~\cite{Bourget2013,Scolan2013,Brouzet2016}, as well as in
inertial waves~\cite{Monsalve2020,Bordes2012} providing a potential route to wave
turbulence~\cite{Davis2020,Brouzet2017,Savaro2020,PanJPO2020}. A special case of TRI is the parametric subharmonic
instability, involving daughter waves with frequencies close to the first subharmonic of the mother
wave and has been well investigated in areas such as plasma physics~\cite{Craik1978,Drake1974} and
oceanic systems~\cite{MacKinnon2013}.

Gravity-capillary waves on the two-dimensional surface of a fluid are well-known in both the
linear regime~\cite{Lamb1932} as well as the nonlinear wave-turbulent
case~\cite{Falcon2007,Falcon2022}. When one of the dimensions of the system is
much smaller than the other, for example in canals, sloshing waves become
apparent~\cite{Lamb1932,Raouf2015}, and lead to longitudinal waves with associated
discrete transverse modes. This results in a countably infinite number of
branches in the dispersion relation, each corresponding to one of the transverse
modes, similar to modes of waveguides~\cite{Laurent2020}.

The sloshing modes can theoretically trigger nonlinear interactions between waves belonging to different branches of the dispersion
relation~\cite{Marchenko1997}. However, the study of the interaction between nonlinear waves of different types (i.e., belonging to separate branches) has not been investigated experimentally so far. Such an unexplored interaction mechanism could
potentially be applicable in various domains such as two-component systems
~\cite{Khusnutdinova2003,Akhatov1995}, atomic lattices~\cite{Pezzi2021}, or plasma
physics~\cite{Tejero2016}. Such multiple branch dispersion relations are also characteristic of
waves in waveguides, for example in solid and soft plates~\cite{Laurent2020}. Besides, it
provides a way to test wave turbulence in media where multiple wave species are
present~\cite{Dias2001} and bears a similarity to interactions between interfacial and free-surface
waves~\cite{Zaleski2020,Issenmann2016}.

Here, we will study the interaction between gravity-capillary waves and the
first sloshing mode. To the best of our knowledge, such three-wave interactions
have not been considered experimentally which is possibly due to the difficulty
of cleanly exciting sloshing modes in typical experiments. The system under
study is a torus of fluid, which has been shown to contain multiple modes of
propagation, including sloshing~\cite{Novkoski2021}, but also easily
demonstrates nonlinear behavior, as shown in the case of Korteweg-de Vries (KdV)
solitons~\cite{Novkoski2022}. Because of its relatively small size
($R\approx 8$\,cm), the torus is easy to manipulate and excite.

The paper is organized as follows. First, we describe in Sec.~\ref{setup} the
experimental setup and the different branches of the dispersion relation.
Section~\ref{experiment} will then present the experimental results related to
the three-wave two-branch interaction. In particular, the sloshing branch can
trigger a triadic resonant instability, generating two gravity-capillary waves,
for which the wave growth rate and phase locking are characterized. The
efficiency of this mechanism is shown to be mediated by a velocity matching
between the two types of propagation modes. Finally, Sec.~\ref{conclusion} draws
the conclusions.

\begin{figure}[t!]
  \includegraphics[width=\columnwidth]{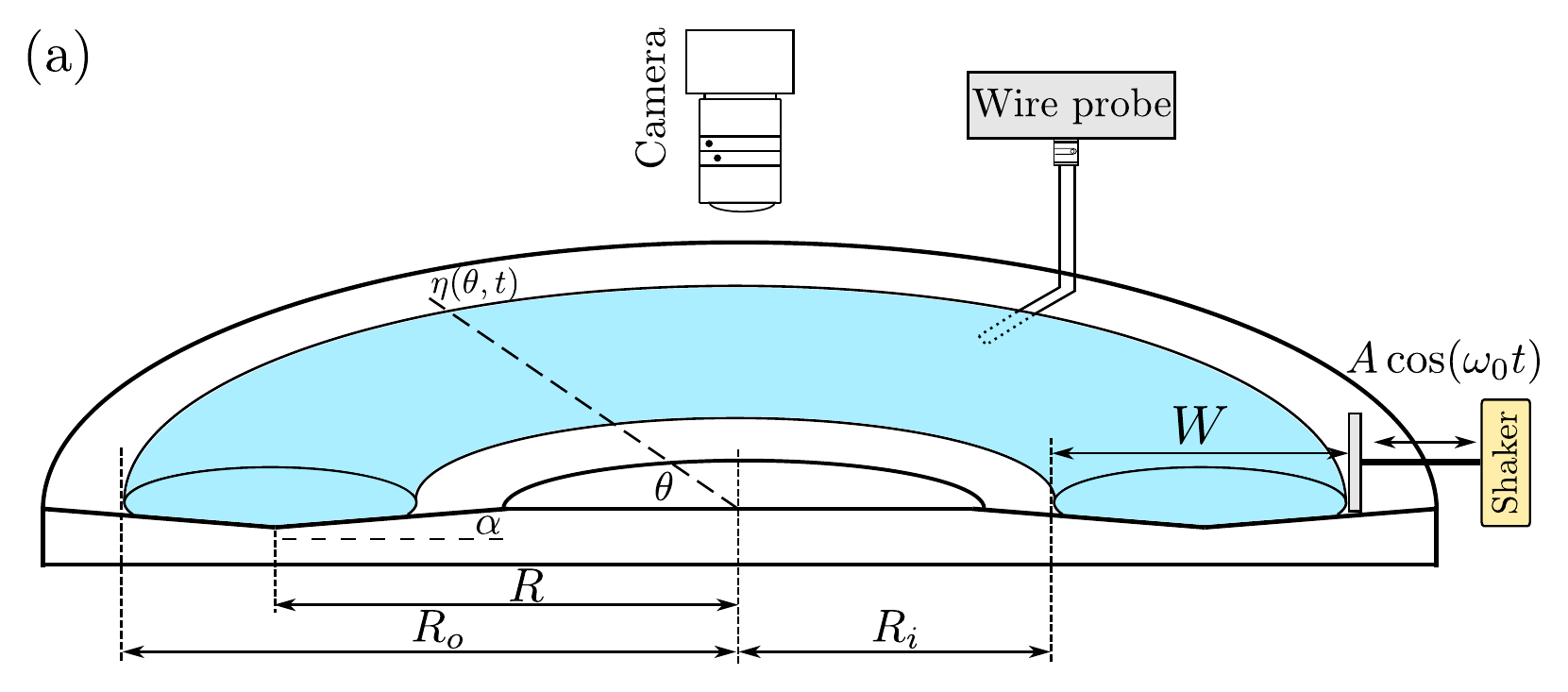}
   \includegraphics[width=\columnwidth]{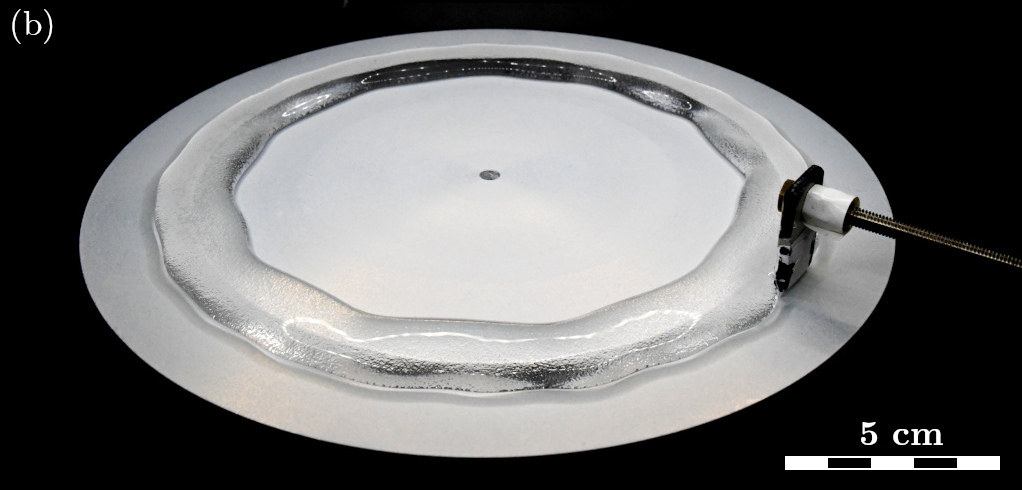}
   \caption{\label{fig:setup}(a) Cross-section of the experimental setup with relevant
     quantities. (b) Side-view of the torus ($R_0=7.85$ cm) under monochromatic excitation,
     $f_0=7.4$ Hz.}
\end{figure}

\section{The torus of fluid}\label{setup}
\subsection{Experimental setup}
The experimental setup used is the same as the one described in Ref.~\cite{Novkoski2021}. The torus of fluid is
formed by depositing distilled water on top of a circular plate that has been coated with a commercial
superhydrophobic treatment. The plate has a triangular groove running along its perimeter, as is
shown in Fig.~\ref{fig:setup}a. The angle of the groove is $\alpha=4.5\degree$. It prevents the
closing of the central hole of the torus due to capillarity.

The waves are created by a Teflon plate connected to an electromagnetic shaker with adjustable sinusoidal
amplitude and frequency typically in the range $7$-$9$ Hz [see Fig.~\ref{fig:setup}b]. The waves
propagate azimuthally on both the inner and outer borders of the torus. The waves also experience
dissipation, which is primarily due to friction of the triple contact line, and not necessarily
viscosity~\cite{Novkoski2022}. The motion of the border of the torus is captured by a camera located directly
above the plate. By using a contour extraction algorithm we obtain the displacement
$\ehh$ of the borders. We will be focusing on the motion of the outer border unless
otherwise mentioned. The second method of detecting the border displacement is through the use of a
custom-made local capacitive wire probe, giving the position of the outer border over time at a
fixed azimuthal point $\theta_0$ with a high temporal resolution ($2$ kHz) [see
Fig.~\ref{fig:setup}a].  The central radius of the groove of the plate is $R=7$ cm, while the torus size is fixed at the
outer border radius $R_o=7.85$ cm. The torus width is then fixed to $W=R_o-R_i=2(R_o-R)=1.7$ cm.

\begin{figure}[t!]
  \includegraphics[width=\columnwidth]{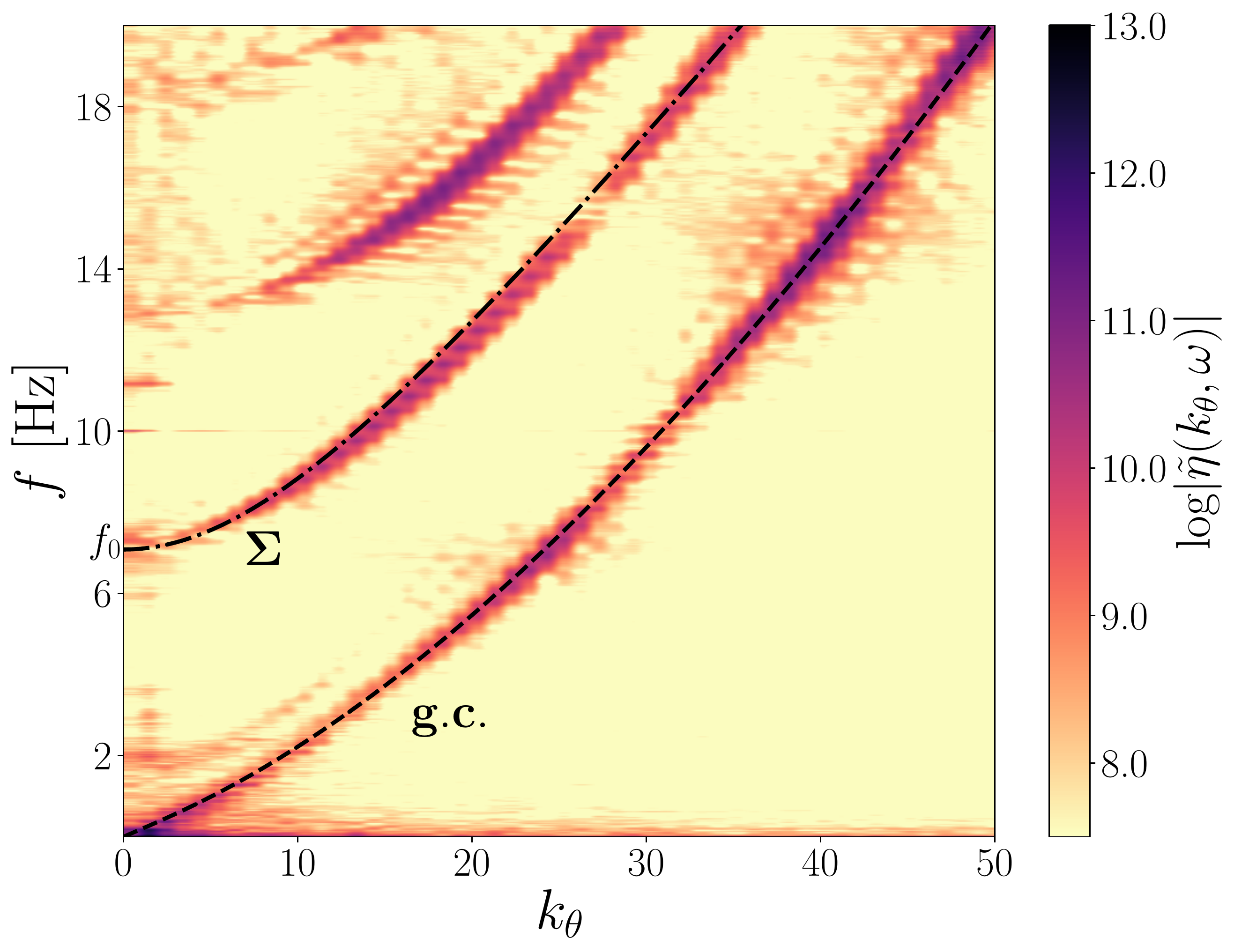}
  \caption{Space-time Fourier spectrum $\tilde{\eta}(\kth,f)$ of the outer border displacement
    $\ehh$ on a torus of fluid, $R_o=7.85$ cm. Forcing: frequency sweep between $0$ and $20$ Hz on
    the outer border. Dashed lines: fit of the gravity-capillary dispersion relation (gc) of
    Eq.~\eqref{eq:dispersion-relation}, and of the sloshing branch, $\Sigma$, given by
    Eq.~\eqref{eq:sloshing}.}
\label{fig:original-spectrum}
\end{figure}


\subsection{Dispersion relation and resonant interaction}
The torus admits several modes of wave propagation such as gravity-capillary azimuthal waves and
sloshing modes~\cite{Novkoski2021}. Using a sweep forcing, the experimental Fourier spectrum
$\tilde{\eta}(\kth,f)$ of the outer border displacement $\ehh$ is shown in
Fig.~\ref{fig:original-spectrum} highlighting the gravity-capillary and sloshing branches.

The dispersion relation of gravity-capillary waves is found to be empirically
well described by~\cite{Novkoski2021}
\begin{align}\label{eq:dispersion-relation}
  \omega^2_{\mathrm{gc}}=\left(g\eff\frac{k_\theta}{R_o}+\frac{\sigma\eff}{\rho}\frac{k_\theta^3}{R_o^3}\right)\tanh{\left(\frac{k_\theta }{R_o}\chi^2\wt\right)} {\rm \,,}
\end{align}
with $\wt=W/2$ the half-width, $\chi=R_o/R$ a measure of curvature, $\rho=1000$ kg$/$m$^{3}$ the density of the fluid, and $\kth$ the angular integer wavenumber, i.e., the discrete mode number which is given as $\kth=kR_o$, with $k$ the dimensional wavenumber of a wave traveling along the torus border. Since the waves are moving on a
slope, they experience an effective gravity which is given by
$g\eff=g\sin\alpha\approx 0.77\,\textup{ms}^{-2}$. The effective surface tension is inferred from
fitting the dispersion relation as $\sigma\eff=55$\,mN$/$m. This low value is due to the channel
geometry and renormalization effects~\cite{LeDoudic2021}.

Alongside the gravity-capillary branch, we consider the first sloshing mode,
also given empirically as~\cite{Novkoski2021}
\begin{align}\label{eq:sloshing}
  \omega_\Sigma^2= \omega_0^2+g\eff\frac{\kth^2}{R}{\rm \,,}
\end{align}
for values of $\kth\lesssim 40$, where $\omega_0$ is the cutoff frequency
at $\kth=0$. This relationship includes only gravity, and for higher frequencies
surface tension needs to be taken into account. In the present work, we will assume
that the relationship is a good approximation at the low wavenumbers we consider here.

\begin{figure}[t!]
  \includegraphics[width=\columnwidth]{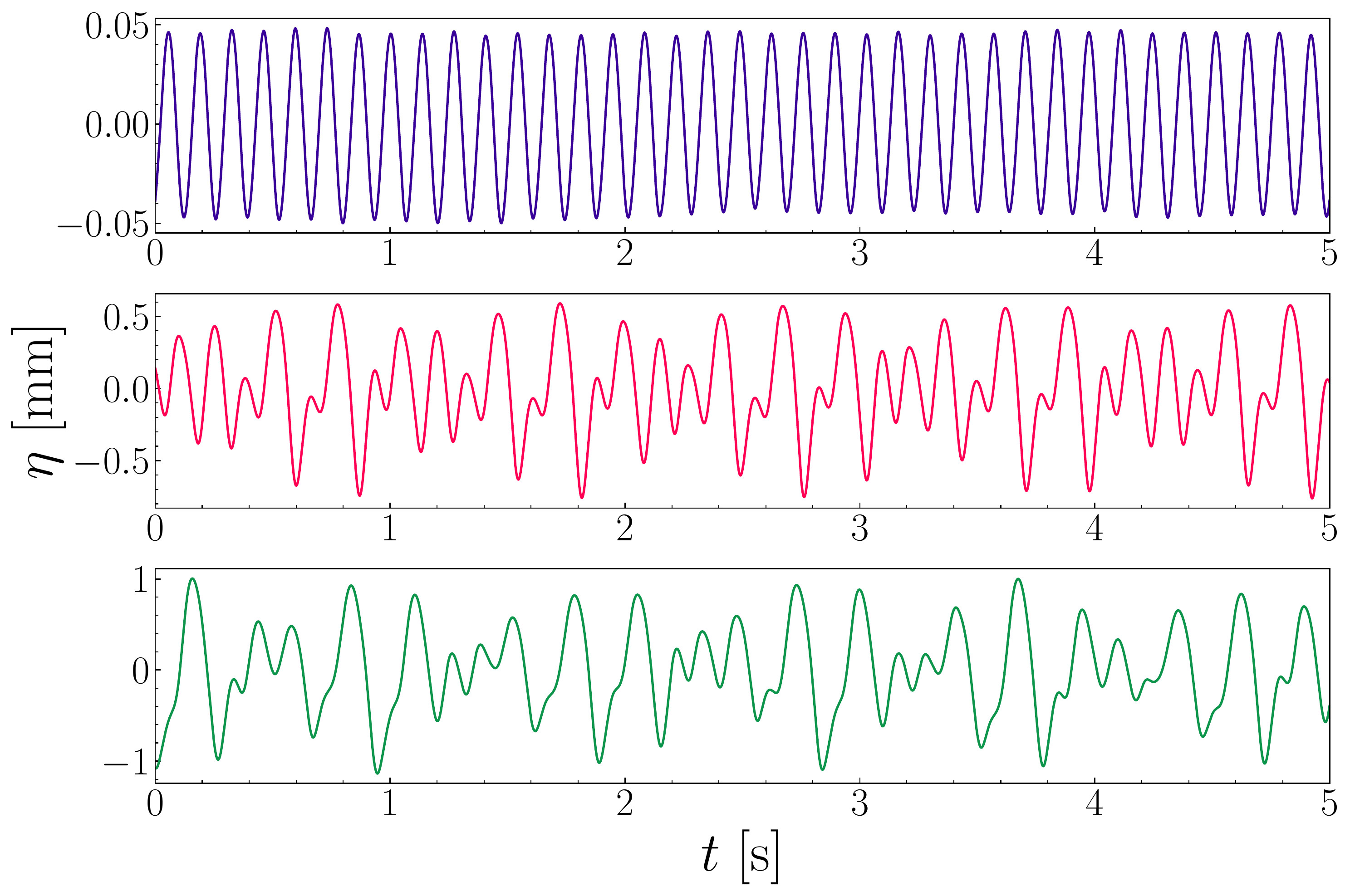}
  \caption{Displacement $\eta(t)$ at a fixed point for three different amplitudes of forcing
    ($f_1=7.4$ Hz) increasing from top to bottom  with the corresponding values of wave steepness
      $\epsilon=0.003$, 0.02, and 0.03. The signal goes from a sine wave into a
    superposition of various subharmonics.}
\label{fig:signals}
\end{figure}

We now turn to the nonlinear interactions between waves. Waves are capable of
exchanging energy through nonlinear resonant interactions if they satisfy
the conservation of both frequency and wavenumber, which in the case of three waves
are
\begin{equation}\label{eq:resonance-generic}
\begin{split}
  k_1 &=k_2+k_3{\rm \,,}\\
  \omega_1 &=\omega_2+\omega_3{\rm \,,}
\end{split}
\end{equation}
with $\omega_i=\omega(|k_i|)$, $\omega(k)$ being the dispersion relation of the
considered system. The above equations can be solved once the dispersion
relation of the waves is provided. In addition, the involved waves do not need
to be of the same type and may belong to different dispersion
branches. Depending on the studied system, solutions of
Eq.~\eqref{eq:resonance-generic} can also give only trivial solutions which do
not lead to an exchange of energy. We will be interested in studying the interaction of
the two different modes mentioned above i.e., between the first two branches in
Fig.~\ref{fig:original-spectrum} (namely the gravity-capillary and first
sloshing branch) and whether they satisfy the conditions given by
Eq.~\eqref{eq:resonance-generic}.

\section{Experimental observations}\label{experiment}
\subsection{Triadic instability}
A monochromatic signal is sent to the shaker at a frequency of $f_1=\omega_1/2\pi$. We measure the
displacement of the outer border $\ett$ at a fixed point for three different amplitudes of
forcing as shown in Fig.~\ref{fig:signals}

At low forcing, $\ett$  very closely resembles a sine wave. By increasing the amplitude, $\ett$
changes significantly, indicating the existence of a critical forcing amplitude and seems to become a
superposition of multiple different frequencies, while still preserving some quasi-periodicity.

\begin{figure}[t!]
  \includegraphics[width=\columnwidth]{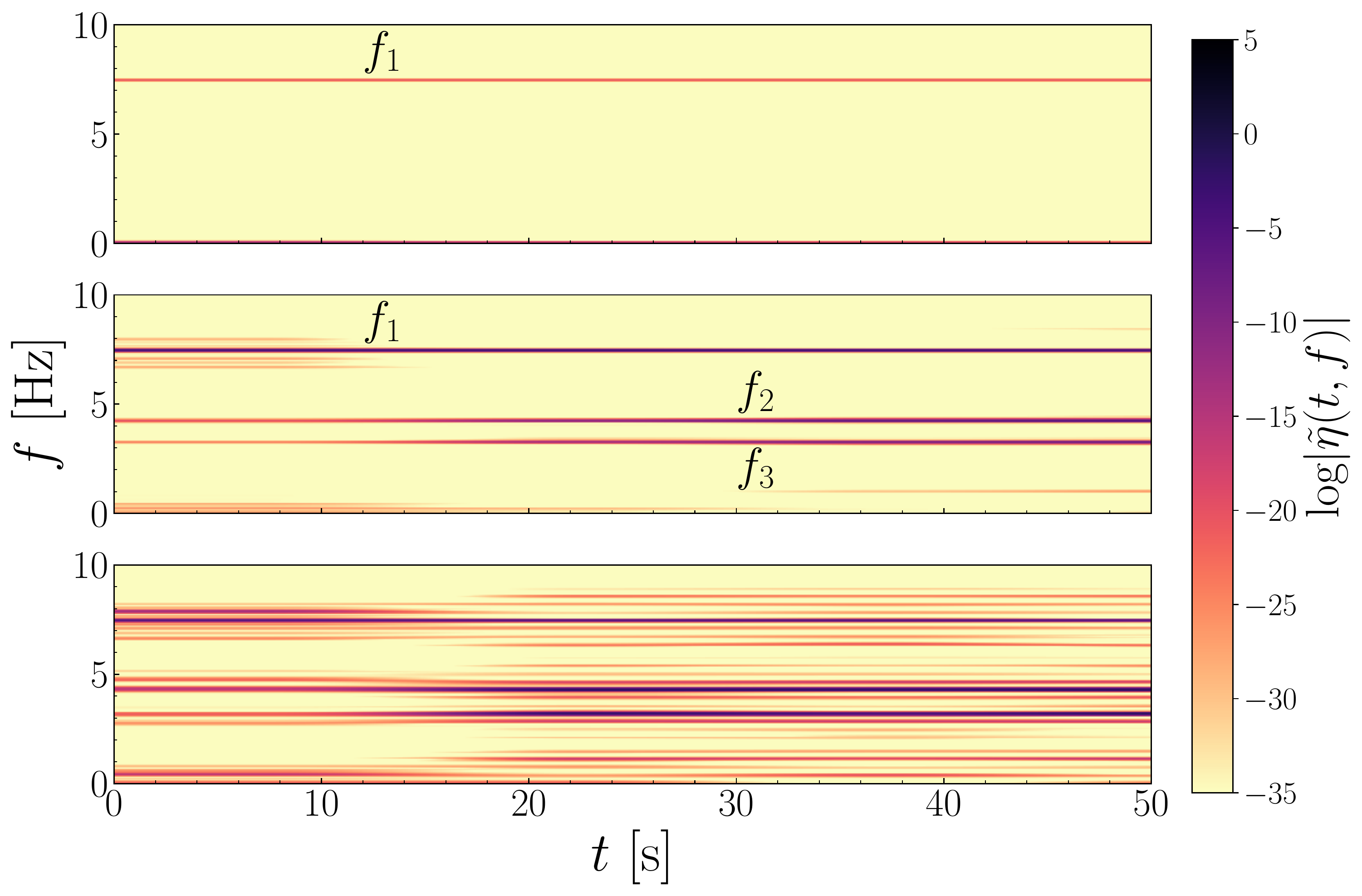}
  \caption{Time-frequency spectrum of the wave amplitudes of
    Fig.~\ref{fig:signals}. For low forcing, a single frequency is present,
    $f_1=7.4$ Hz (top), but for a high enough forcing, two frequencies appear at
    $f_2=4.2$ Hz and $f_3=3.2$ Hz (middle). Further increase of the forcing
    generates an ensemble of different modes (bottom).}
\label{fig:spectrogram}
\end{figure}

We compute the time-frequency spectrum of $\ett$ (also called spectrograms) to distinguish the
frequency components contained in the signal, as shown in Fig.~\ref{fig:spectrogram} for the three
forcing amplitudes. For a low forcing, a single frequency is found, corresponding precisely to the
forcing one, $f_1$. As the forcing is increased, two additional subharmonic frequencies appear,
neither of which is located at $f_1/2$. This behavior is characteristic of the triadic resonant
instability, where, by forcing the system at a given \textit{pump frequency} $f_1$, two subharmonic
waves, $f_2$ and $f_3$ are pumped up from zero amplitude and thus begin to deplete the pump. It is
worth noting that the sum of these two new frequencies, $f_2$ and $f_3$ equals $f_1$. We also see
that a typical time (of the order of $10$ s) is necessary for these waves to be established in the
spectrum. As the amplitude of forcing is increased further, additional frequencies besides the first pair
become visible but take more time to appear, and they do so after the original pair is established.

\subsection{Resonance conditions}
\begin{figure}[t!]
  \includegraphics[width=\columnwidth]{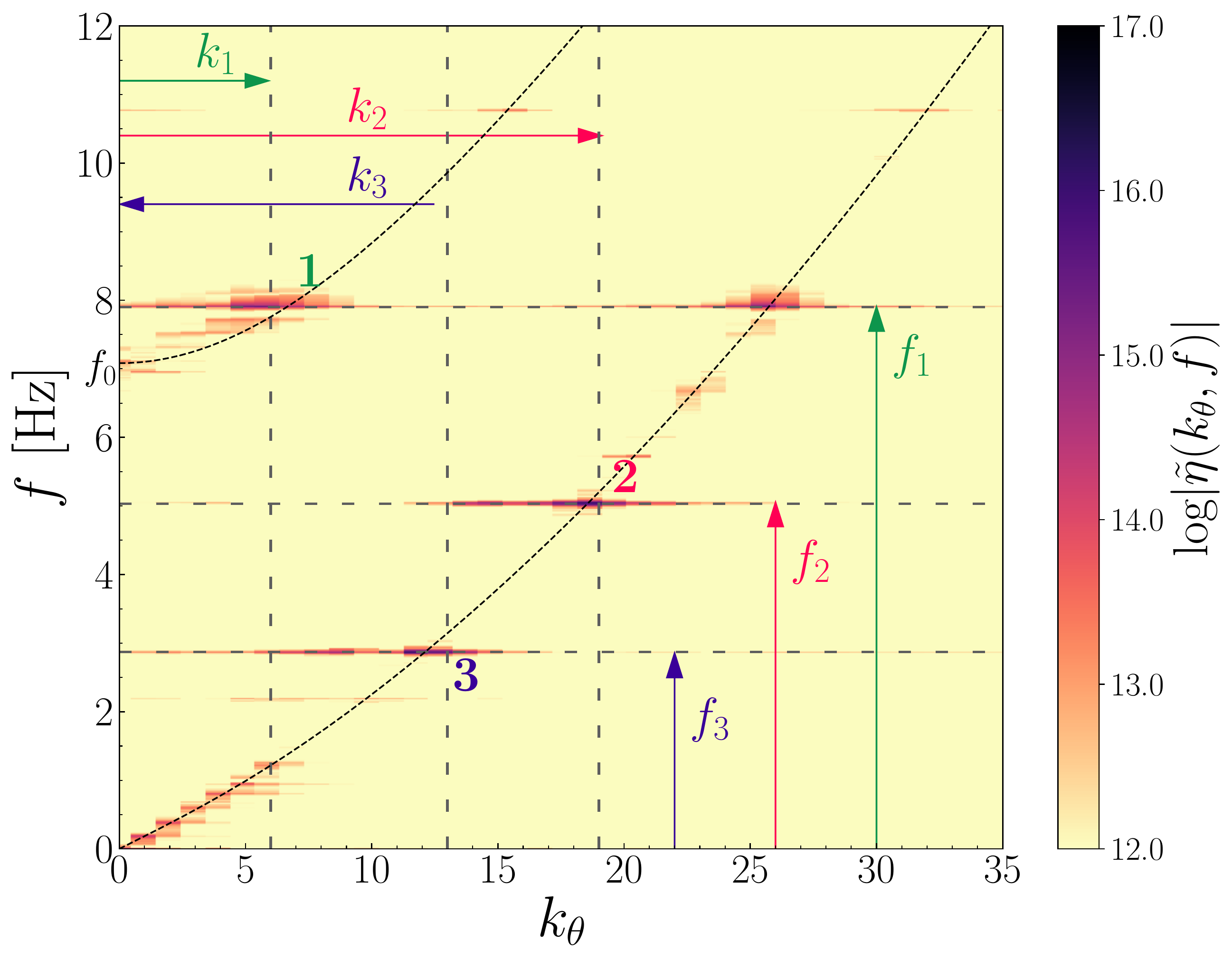}
  \caption{Fourier spectrum $\tilde{\eta}(\kth,f)$ of the torus outer border
    displacement $\ehh$.  Monochromatic forcing at $f_1=7.9$ Hz. Points
    $(k_2,f_2)$ and $(k_3,f_3)$ lie on the gravity-capillary branch whereas,
    point $(k_1,f_1)$ lies on the sloshing branch where it is forced. All three
    points verify the resonance conditions in both frequency and wavenumber (see
    arrows). $f_0$ is the cutoff frequency of the sloshing branch.}
\label{fig:spectrum}
\end{figure}

To verify that the signals we observe are due to a resonant three-wave interaction, we first
consider the spatiotemporal signal of the torus outer border $\eta(\theta,t)$. We then compute the
corresponding space and time Fourier transform $\tilde{\eta}(\kth,\omega)$ as shown in
Fig.~\ref{fig:spectrum}. It gives us not only the frequency but also wavenumber information of the
waves present in the system. Figure~\ref{fig:spectrum} shows that the pumping frequency $f_1$
excited the sloshing branch, and the corresponding part on the gravity-capillary branch, but also
two lower frequency points, $f_2$ and $f_3$ on the gravity-capillary branch. Note that the discreteness in $\kth$ is due to the torus finite size, whereas the one in $f$  corresponds to the inverse of the total measurement time.

\begin{figure}[t!]
  \includegraphics[width=\columnwidth]{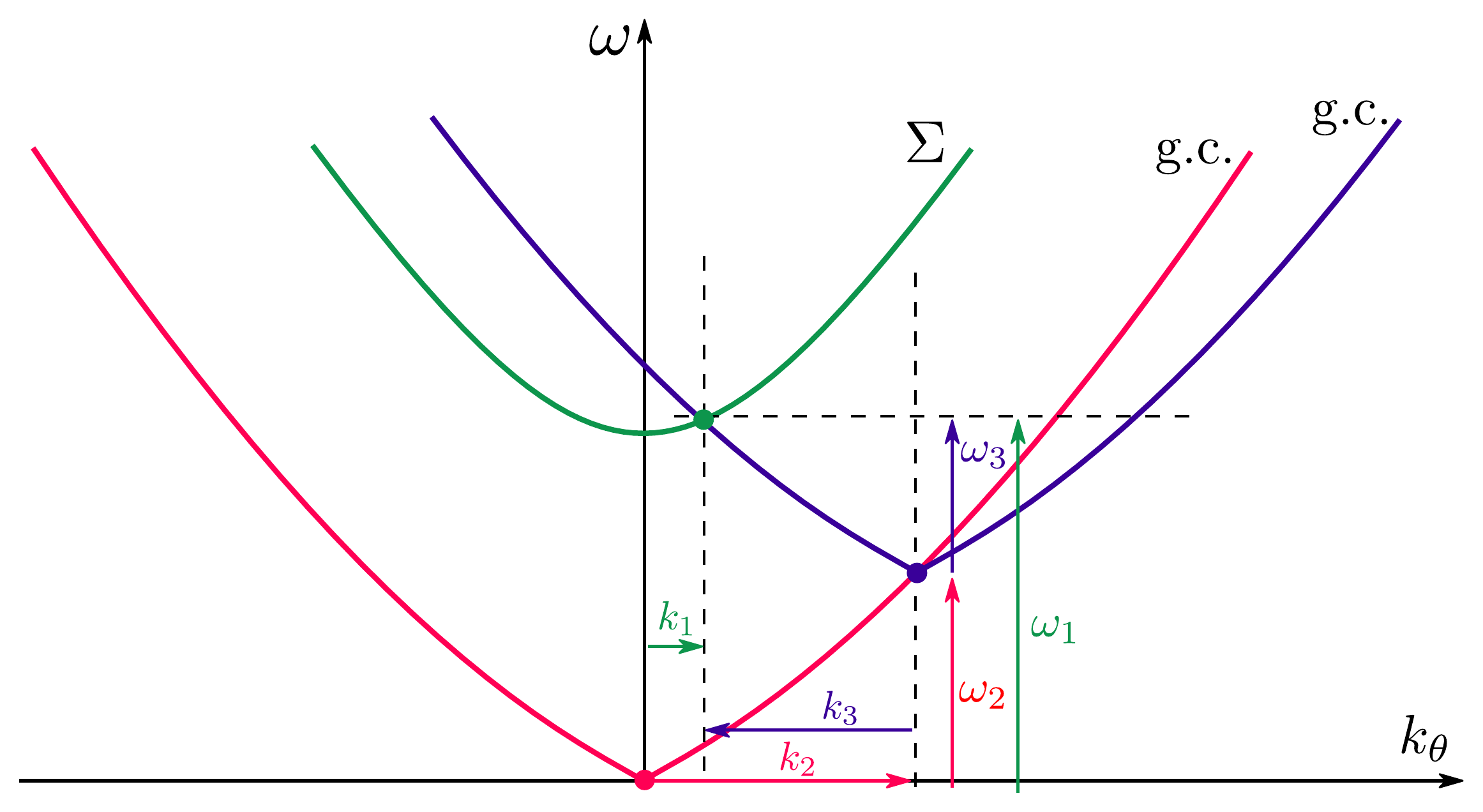}
  \caption{Graphical solution of the resonance conditions of Eq.~\eqref{eq:resonance} involving a
    sloshing wave decomposing into two gravity-capillary waves using the corresponding dispersion
    relations. One of the waves has to be counter-propagating to verify Eq.~\eqref{eq:resonance}.}
\label{fig:demo}
\end{figure}

Thus, one has to consider the following resonant conditions
\begin{align}\label{eq:resonance}
  \begin{split}
  \omega_1^{\Sigma}&=\omega_2^{{\mathrm{gc}}}+\omega_3^{{\mathrm{gc}}}{\rm \,,} \\
  \kth\left(\omega_1^\Sigma\right)&=\kth\left(\omega_2^{{\mathrm{gc}}}\right)+\kth\left(\omega_3^{{\mathrm{gc}}}\right){\rm \,,}
  \end{split}
\end{align}
which can be solved graphically in the $(\kth,\omega)$ plane as demonstrated in
Fig.~\ref{fig:demo}. Experimentally we find the resonance conditions in both
wavenumber and frequency to be verified by the points 1, 2, and 3 displayed in
Fig.~\ref{fig:spectrum}. Since the forcing frequency is known at all times,
i.e., $\omega_1=\omega_1^\Sigma$, we solve exactly the above
Eq.~\eqref{eq:resonance}, using the two branches of the dispersion relation of
Eqs.~\eqref{eq:dispersion-relation} and \eqref{eq:sloshing}, leading to a system
of four equations and four unknowns. Since no analytic solution exists, we look
for the two unknown daughter frequencies $f_2$ and $f_3$ numerically. It is also
important to note that one of the daughter waves will always have a negative
wavenumber, i.e., it will be counterpropagating with respect to the other two
waves. If exclusively three interacting gravity-capillary waves are taken into
account (i.e., no sloshing), no nontrivial solution to the above equations
exists far from the capillary-gravity transition~\cite{NazarenkoBook}.

\begin{figure}[b!]
  \includegraphics[width=\columnwidth]{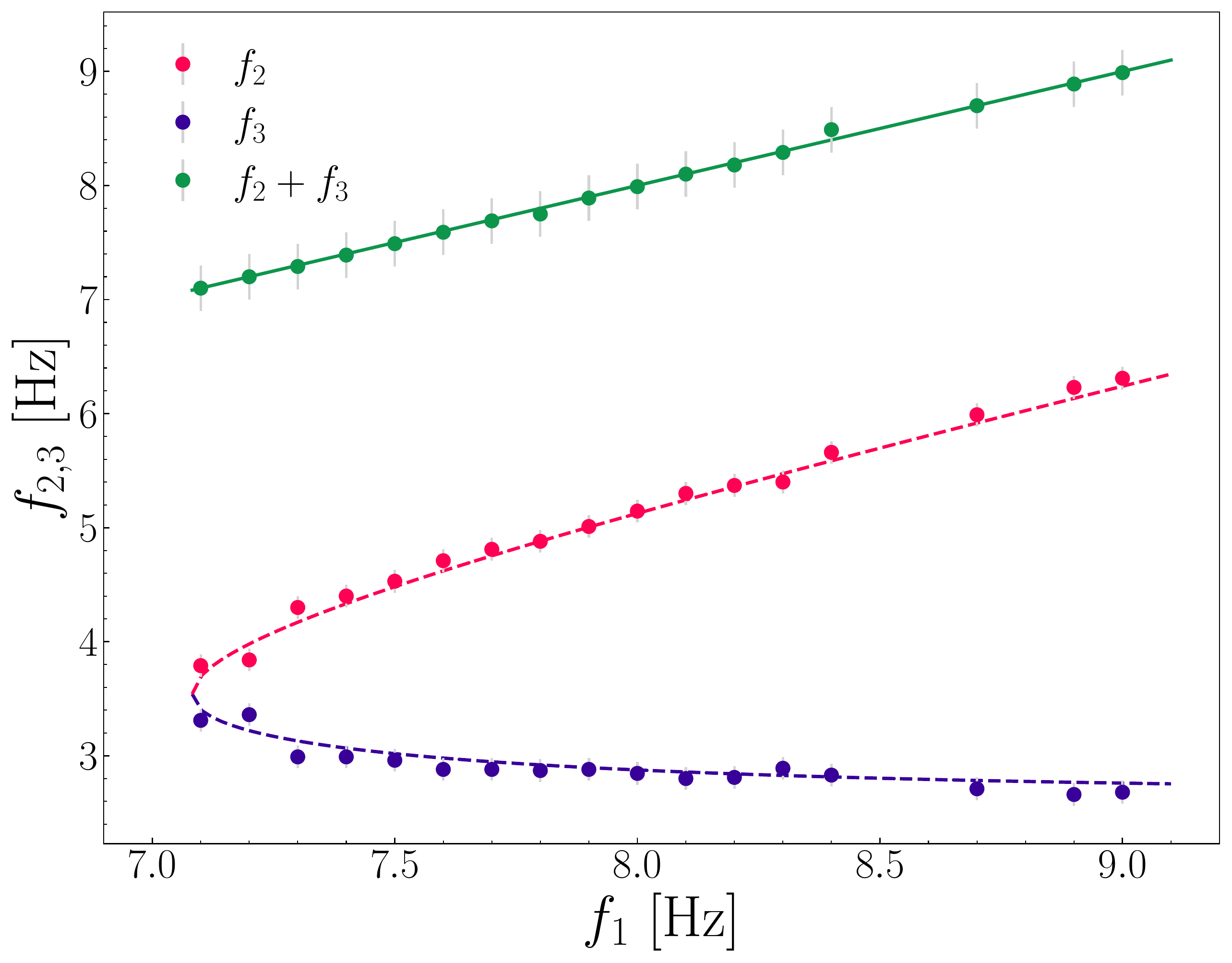}
  \caption{Measured values (dots) of the daughter frequencies $f_2$ (red) and $f_3$ (blue) for
    different mother frequencies $f_1$. The dashed lines are theoretical values obtained numerically
    through the resonance conditions of Eq.~\eqref{eq:resonance}. Solid line of slope 1 corresponds
    to $f_1$ and green dots indicate the sum $f_2+f_3$.}
\label{fig:resonance}
\end{figure}

Due to the periodicity of the system and its finite size, the dispersion
relation of the torus is necessarily discrete~\cite{Novkoski2021}. If we denote
by $\Delta_\omega\equiv \omega(\kth+1)-\omega(\kth)$ the frequency gap between two adjacent discrete wavenumber $\kth$
and by $\Gamma_\omega$ the frequency nonlinear broadening of the dispersion relation, we can experimentally estimate whether
discrete effects are to be taken into account ($\Gamma_\omega/\Delta_\omega\ll 1$)
or if the system is in a kinetic regime ($\Gamma_\omega/\Delta_\omega\gg 1$)~\cite{Lvov2010}. Indeed, we find
approximately that $\Gamma_\omega/\Delta_\omega\in [3,6]$ in our experiment, hence we
can consider the dispersion relation continuous.

The two daughter frequencies are now measured for different values of the mother frequency $f_1$ to
confirm that the resonance conditions are well satisfied. Both the numerical solution of
Eq.~\eqref{eq:resonance} (dashed lines), and the experimentally found values (dots) of $f_1$ and
$f_2$ are in very good agreement as shown in Fig.~\ref{fig:resonance}. As we can see the frequencies
satisfy the conditions extremely well, and not only confirm the frequency matching condition but
also wavenumber conservation since this is implicitly included when solving the resonance conditions
in Eq.~\eqref{eq:resonance}. This confirms that the system is experiencing a resonant three-wave
(two-branch) interaction. We note that below the cutoff frequency $\omega_0$ of the sloshing branch
($f_0<7.1$ Hz - see Figs.~\ref{fig:original-spectrum} and \ref{fig:spectrum}), no
solution occurs in Fig.~\ref{fig:resonance}. Indeed, forcing below $f_0$ does not
lead experimentally to the appearance of nonlinear resonant interactions.

\begin{figure}[t!]
  \includegraphics[width=\columnwidth]{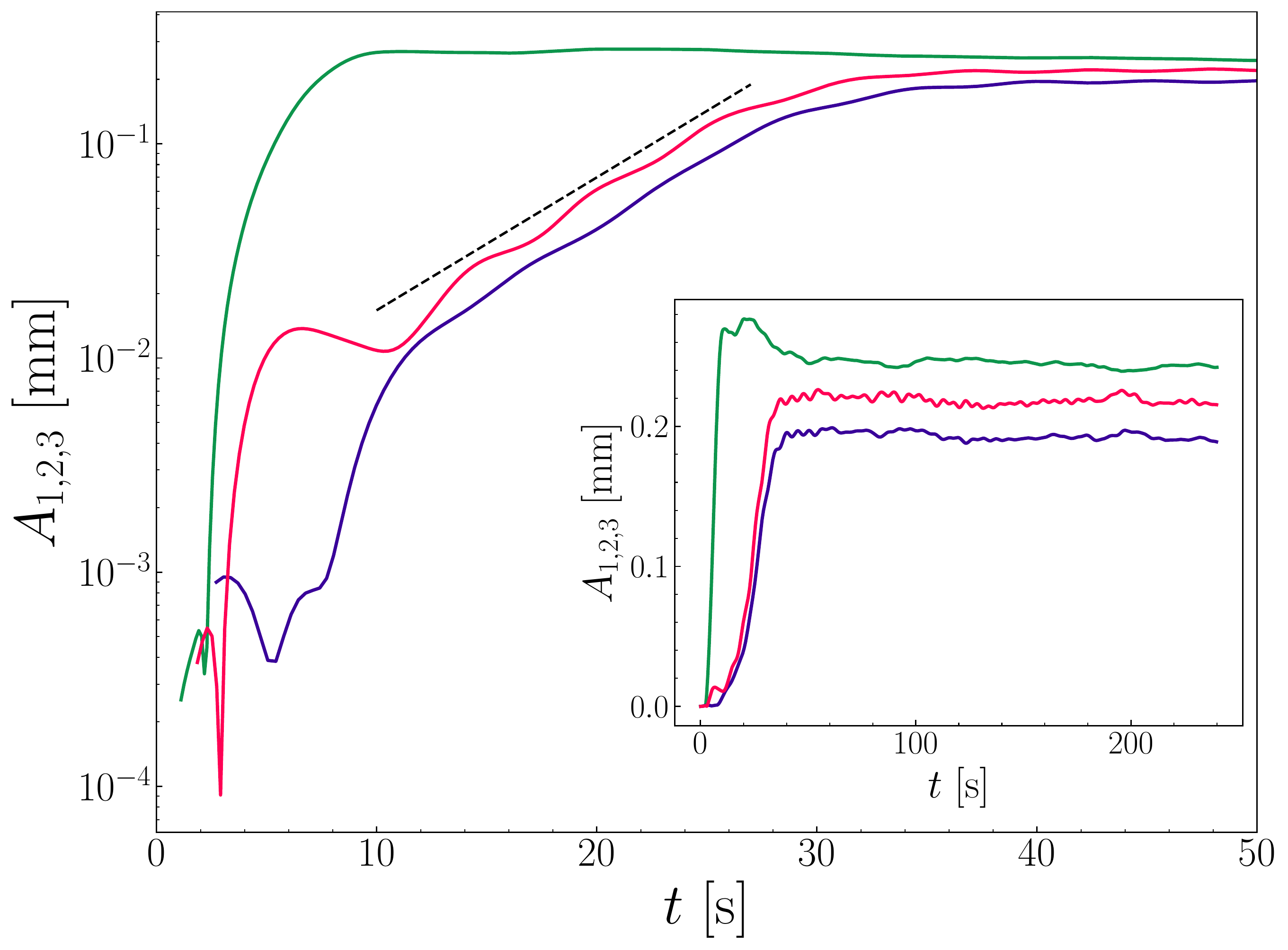}
  \caption{Semilog plot of amplitudes $A_i$ as a function of time of the mother wave (green,
    $f_1=7.4$ Hz), and the two daughter waves (in red and blue) measured using the Hilbert
    transform. We can observe that at around $12$ s the two daughters start growing exponentially as
    $e^{t/\tau}$ with $\tau=7$ s (dashed line). Inset: same in linear scale for the whole duration
    of the experiment, $T=240$ s. We can see how initially, as the daughter waves grow, the mother
    wave has to lose energy.}
\label{fig:hilbert-amplitudes}
\end{figure}

\subsection{Wave amplitude growth}
Three-wave interactions are usually described using amplitude equations. We now consider the
amplitudes of each wave at frequency $f_i$, which are experimentally accessible through the use of
the Hilbert transform of $\ett$~\cite{Bonnefoy2016}. This is done by first using a bandpass filter
around the frequency of interest, onto which the Hilbert transform is then applied. This procedure
then yields both the wave amplitude at frequency $f_i$ but also its phase $\varphi_i$.

We focus first on the displacement $\ett$ forced at $f_1=7.4$ Hz from which we extract the three
amplitudes. As we saw in Fig.~\ref{fig:spectrogram} (middle), some typical time is needed for the
transfer of energy from the mother wave $f_1$ into the daughters $f_2$ and $f_3$. The temporal
evolutions of the amplitudes of all three waves are shown in Fig.~ \ref{fig:hilbert-amplitudes}. Once
the mother wave is established, ($t<3$ s) it then increases rapidly up to a stationary
out-of-equilibrium state ($t>10$ s). The growth is indeed balanced by dissipation when it begins
pumping the daughter waves which grow exponentially, ($12<t<25$ s). Eventually, all three reach a
stationary state ($t>35$ s). The exponential growth of the two daughter waves indicates that they
undergo an instability. In addition, the mother wave reaches an initially higher amplitude
which then decreases to the steady one, since it transfers energy to the two daughter waves through
the instability. It is important to note that the three amplitudes are of the same order of
magnitude, and the mother wave cannot be considered to be much stronger than the daughter waves as it
is usual in three-wave resonant interactions~\cite{Haudin2016,Bonnefoy2016,Craik1978}.

\begin{figure}[t!]
  \includegraphics[width=\columnwidth]{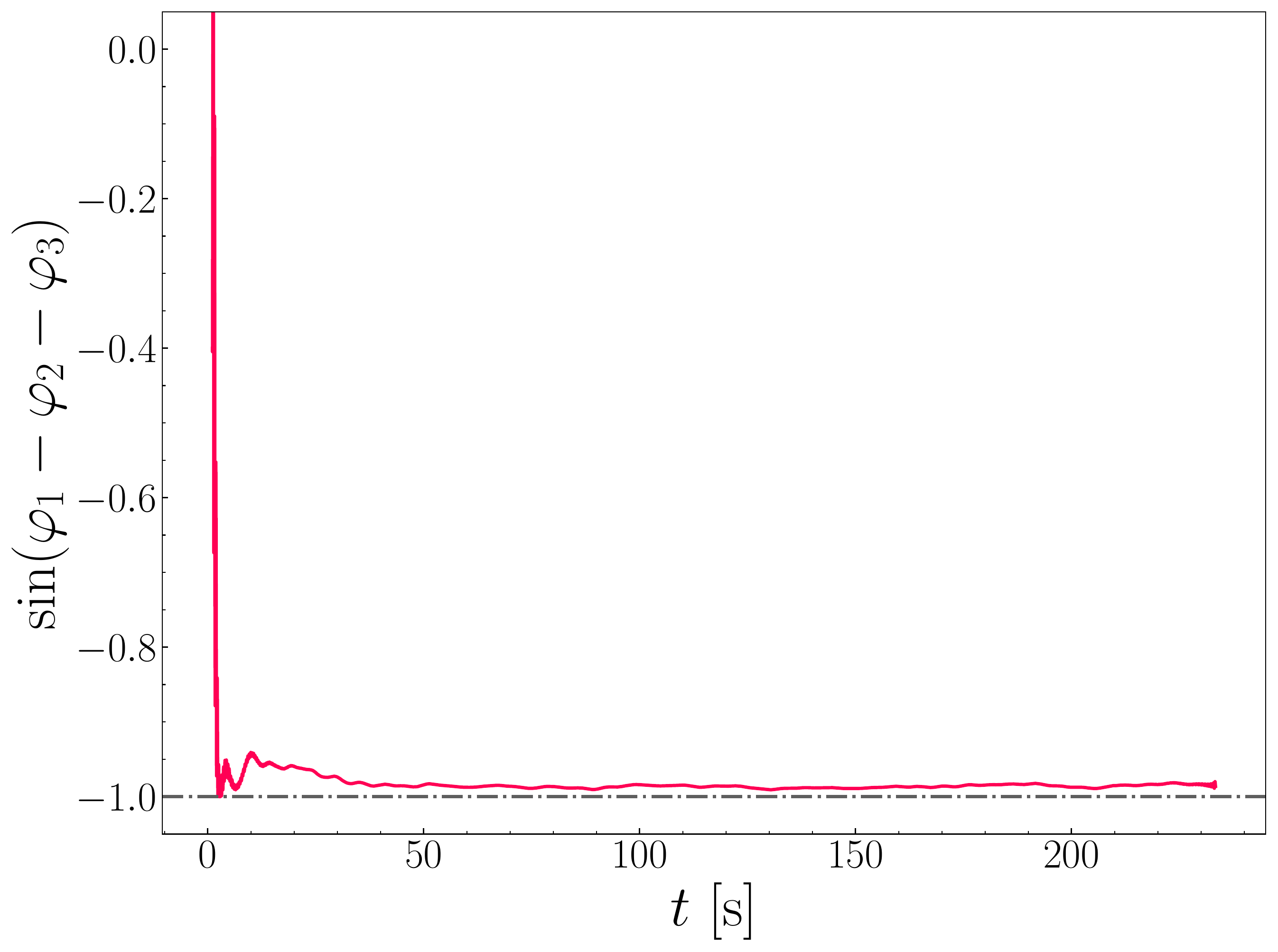}
  \caption{Temporal evolution of the sine of the total phase $\varphi=\varphi_1-\varphi_2-\varphi_3$
    of the three waves obtained using the argument of the Hilbert transform. The total phase
    $\varphi$ is found to be locked to a value close to $-\pi/2$. Same forcing as in
    Fig.~\ref{fig:hilbert-amplitudes}.}
\label{fig:hilbert-phase}
\end{figure}

\subsection{Phase locking}
The phase of each wave reads $\varphi_j(\theta,t)=\kth\theta-\omega t+\phi_{j}$
and in general, it depends on time, $\phi_j$ being an initial arbitrary
constant. Conversely, when Eq.~\eqref{eq:resonance} is satisfied, the interaction
phase defined as $\Phi=\varphi_1-\varphi_2-\varphi_3$ remains constant (i.e.,
$\phi_1-\phi_2-\phi_3=\mathrm{const.}$), thus making the three waves
phase-locked. Experimentally, $\varphi_i$ is measured by the argument of the
Hilbert transform of $\ett$ in the stationary regime.

To avoid possible phase jumps, we plot the sine of the total phase $\Phi$ in
Fig.~\ref{fig:hilbert-phase}. Once the stationary regime is reached, the total
phase remains constant over time. The three waves are thus phase-locked around
$\Phi\simeq-\pi/2$ as expected theoretically for a three-wave resonant
mechanism (see Sec.~\ref{AmpEq}).

\subsection{Amplitude equations}\label{AmpEq}
We now consider the three-wave amplitude equations in the case of resonant
interaction~\cite{Craik1986,Marchenko1997} for a physical description of this instability
\begin{align}
  \dot{A}_1=iI_{23}A_2A_3{\rm \,,} \\
  \dot{A}_2=iI_{13}A_1A_3^*{\rm \,,} \\
  \dot{A}_3=iI_{12}A_1A_2^*{\rm \,,} 
\end{align}
with $A_i$ the complex wave amplitude and $I_{i,i+1}$ are the unknown positive
interaction coefficients. Note that the latter are known for gravity-capillary
wave interaction involving no sloshing~\cite{Simmons1969}. We will not approach
the full problem of the above equations, which constitute an integrable
system~\cite{Kaup1976}. We instead focus only on the case where the pump-wave
has a fixed amplitude $A_1$ (the so-called pump-wave approximation). Indeed, we saw
experimentally that the stationary regime of the mother wave is established before
the one of two daughter waves. For completeness, we include damping as well, leading
to
\begin{align}
  \dot{A_2}=iI_{13}A_1A_3^* -\alpha_2A_2{\rm \,,} \label{eq:amp1} \\
  \dot{A_3}=iI_{12}A_1A_2^*-\alpha_3A_3{\rm \,,}  \label{eq:amp2}
\end{align}
with $\alpha_j$ the temporal damping rate of wave $j$. Inserting Eq.~\eqref{eq:amp1} into
Eq.~\eqref{eq:amp2} leads to
\begin{align}
  \ddot{A_3}=I_{13}I_{12}A_3|A_1|^2 -(\alpha_2+\alpha_3)\dot{A}_3-\alpha_2\alpha_3A_3{\rm \,,}
\end{align}
which has a solution of the form
\begin{align}
  A_3=a_+e^{\sigma_+t}+a_-e^{\sigma_-t}{\rm \,,} 
\end{align}
with $a_\pm$ depending on the initial conditions and the growth rate obeying
\begin{align}
  \sigma_{\pm}=-\frac{\alpha_2+\alpha_3}{2}\pm\sqrt{I_{12}I_{13}|A_1|^2+\frac{(\alpha_2-\alpha_3)^2}{4}}{\rm \,.}
\end{align}
Thus an instability (i.e., $\sigma_+>0$) can be observed provided the mother
amplitude overcomes a threshold due to dissipation. The daughter waves then grow
exponentially, as observed experimentally. This means that if at time $t=0$,
only wave $1$ has a finite amplitude, the other two waves, which are
infinitesimal in magnitude, will be pumped up exponentially, and eventually be
bounded by damping. The exact values of the interaction coefficients would
follow from a weak nonlinear expansion of equations of motion, which for the
case of the torus in this experimental geometry are so far unknown.

As for the phases of the waves, denoting $A_j=a_je^{i\varphi_j}$ yields
the equation for the temporal evolution of the total phase~\cite{Craik1986}
\begin{align}
  \dot{\Phi}=a_1a_2a_3\left(\frac{I_{23}}{a_1^2}-\frac{I_{13}}{a_2^2}-\frac{I_{12}}{a_3^2}\right)\cos\Phi=\beta\cos\Phi{\rm \,.}
\end{align}
If we consider that in the final stationary state all three amplitudes are constant, one finds a
solution of the form
\begin{align}
  \Phi=2\arctan\left[\tanh\left(\frac{\beta (t_0+t)}{2}\right)\right]{\rm \,,}
\end{align}
which at large $t$ leads to $\varphi=\textup{sgn}(\beta)\pi/2$. Depending on the sign of $\beta$,
i.e., the values of the interaction coefficients and amplitudes, the sign of the total phase will be
differently determined, which we find experimentally to be $-\pi/2$. We find that the interaction
phase $\Phi$ does not change with the frequency of the mother wave in the experiment.

\begin{figure}[t!]
  \includegraphics[width=\columnwidth]{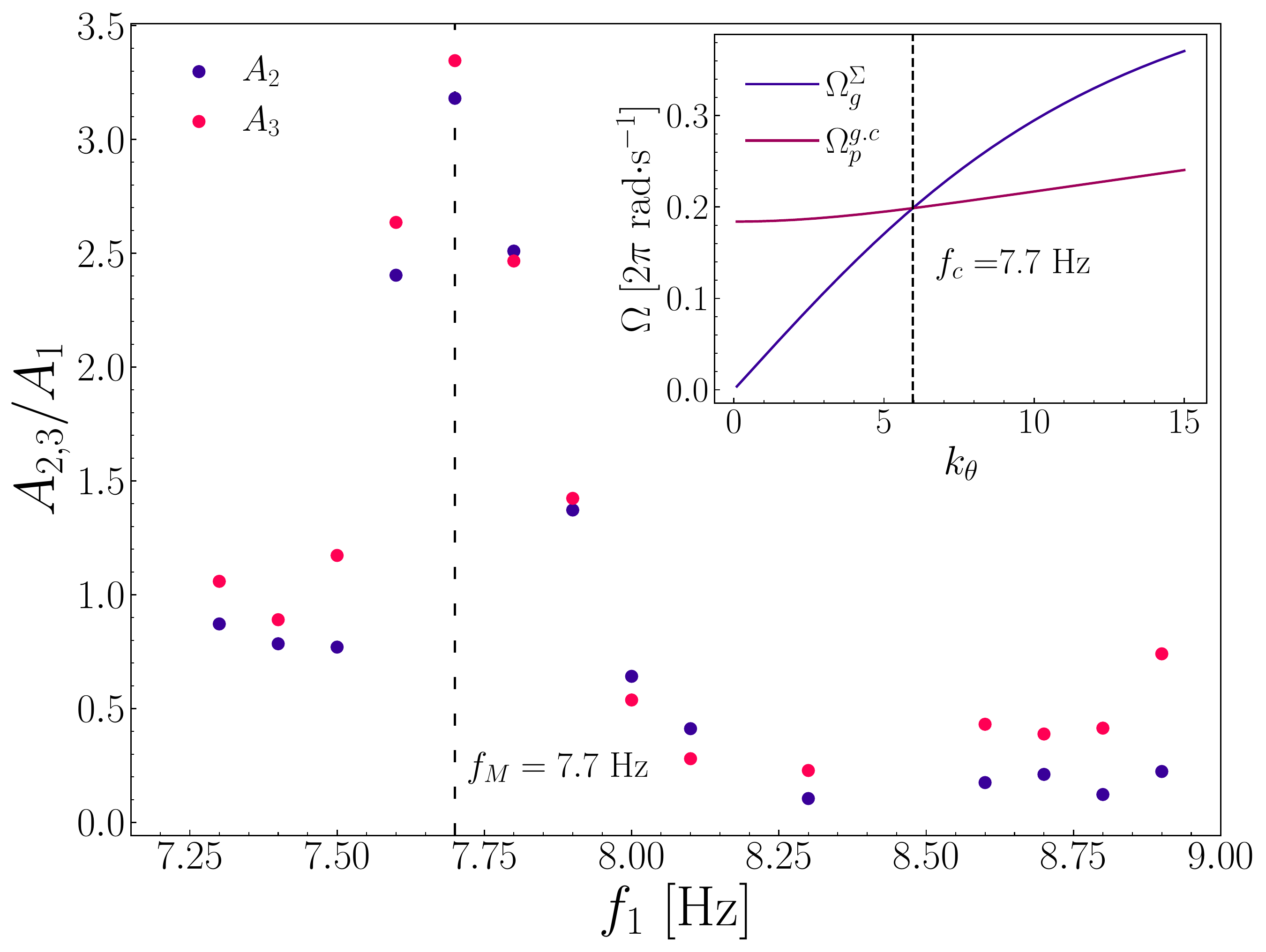}
  \caption{Normalized amplitudes of the two daughter waves for different
    frequencies of the mother wave $f_1$. We clearly observe a peak at around
    $f_M\approx7.7$ Hz, where the daughter waves are three times larger than the
    mother wave in amplitude. Inset: Predicted group velocity of the sloshing
    mode (blue) and phase velocity of gravity-capillary mode (red), intersecting
    at $k_c=6$ corresponding to the peak frequency $f_M=7.7$ Hz.}
\label{fig:frequency-sweep}
\end{figure}

\subsection{Maximal energy transfer by velocity matching}
We now turn to the dependence of the daughter amplitudes on the frequency $f_1$
of the mother wave. The daughter wave amplitudes normalized by the mother wave
$A_{2,3}/A_1$ are shown as a function of $f_1$ in
Fig.~\ref{fig:frequency-sweep}. The two daughter waves appear to follow the same
relation and experience a maximal relative amplitude, making their amplitudes
significantly larger than the mother wave. The plot is strongly reminiscent of the
resonance curve of a driven harmonic oscillator. The peak of this curve,
experimentally found to be at $f_{M}=7.7$~Hz, has to be located at a frequency
that depends only on the system properties. We find it to be close to the
frequency $\omega_\Sigma(k_c)$ where $k_c=6$ is the wavenumber at which the
group velocity, $\Omega^{\Sigma}_g=\textup{d}\omega_\Sigma/\textup{d}\kth$, of
the sloshing branch and the phase velocity of the gravity-capillary, $\Omega^{\mathrm{gc}}_p=\omega_{\mathrm{gc}}/\kth$, intersect,
numerically found to be $f_c\approx7.7$~Hz, shown in the inset of Fig.
\ref{fig:frequency-sweep}.

The energy transfer is thus most efficient when a velocity matching occurs
between the group velocity of the sloshing mode and the phase velocity of the
gravity-capillary mode. Such an atypical velocity matching involving group and
phase velocities differs from the usual phase-phase velocity
matching~\cite{Fedorov1998}, but has been considered theoretically for surface
and internal waves~\cite{Kawahara1975,Taklo2020}. The energy transfer is thus
found to be maximal when the carrier of a sloshing wavepacket has the same
velocity as a gravity-capillary monochromatic wave for an identical wavenumber. Note that the
efficiency of the wave interaction is thus related to the velocity matching, whereas
the triadic interaction is the transfer mechanism.

Sloshing branches have previously been modeled, in the linear case, using systems of
oscillators~\cite{Ibrahim2001}, while their nonlinear interaction remains more complicated. A model
of the branch interaction would have to resemble the driven harmonic oscillator whose amplitude
depends on $\omega_c^2-\omega_3^2$, similar to that found in~\cite{Cazaubiel2019}, where
$\omega_c=\omega_\Sigma(k_c)$ is the frequency at which $\Omega_p^{\mathrm{gc}}=\Omega_g^\Sigma$.


\begin{figure}[t!]
  \includegraphics[width=\columnwidth]{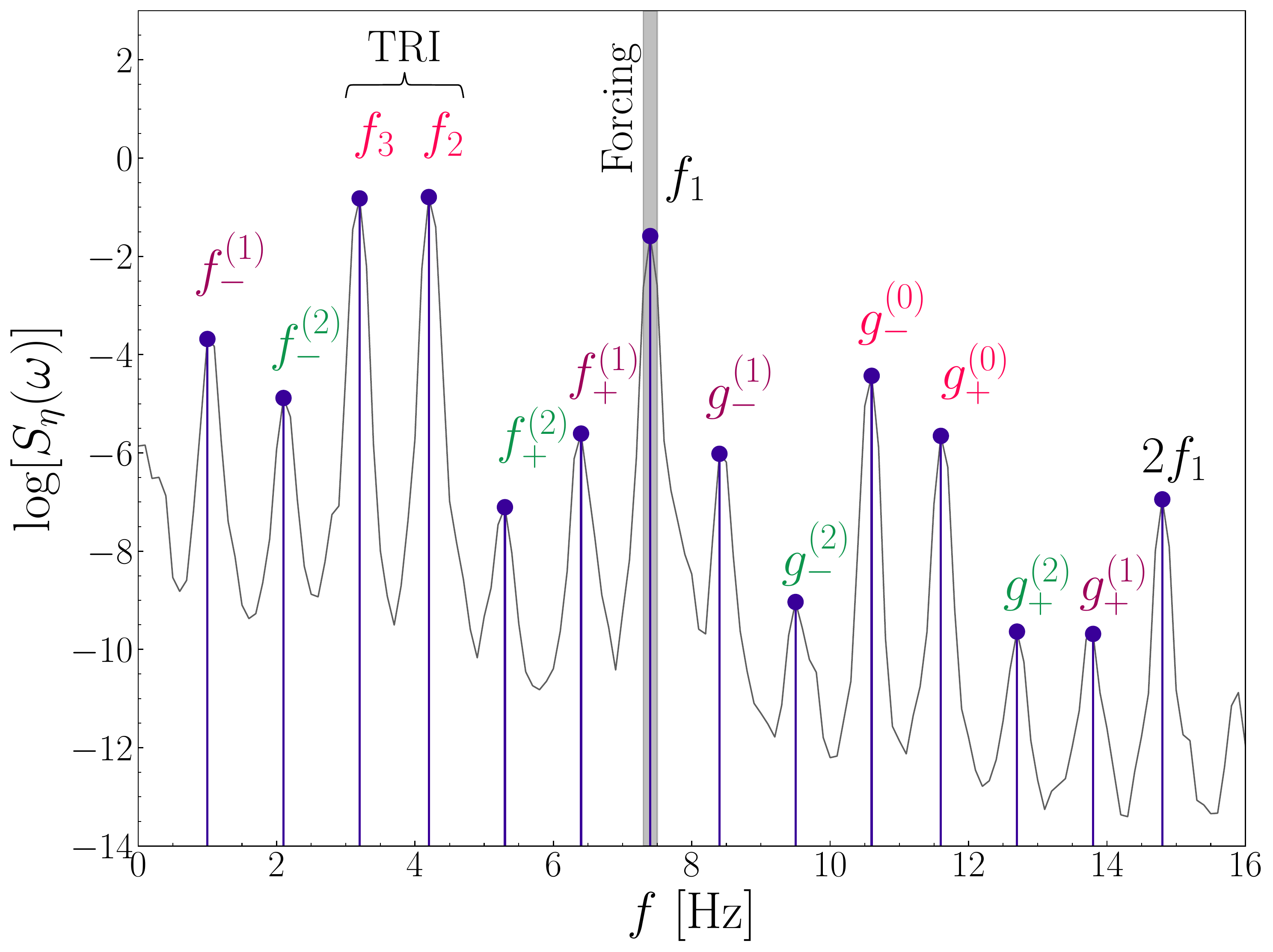}
  \caption{Power spectrum of a signal forced at $f_1=7.4$ Hz. The forcing is
    strong enough to excite additional couples besides the primary three-wave
    pairs, $f_1^{(1)}$ and $f_2^{(1)}$. The first two daughters, continue to
    generate firstly the subharmonic secondary waves of Eq.~\eqref{eq:couples},
    which then go on to create tertiary waves through interactions with the
    mother wave at $f_1$.}
\label{fig:combinations}
\end{figure}

\begin{figure}[t!]
  \includegraphics[width=\columnwidth]{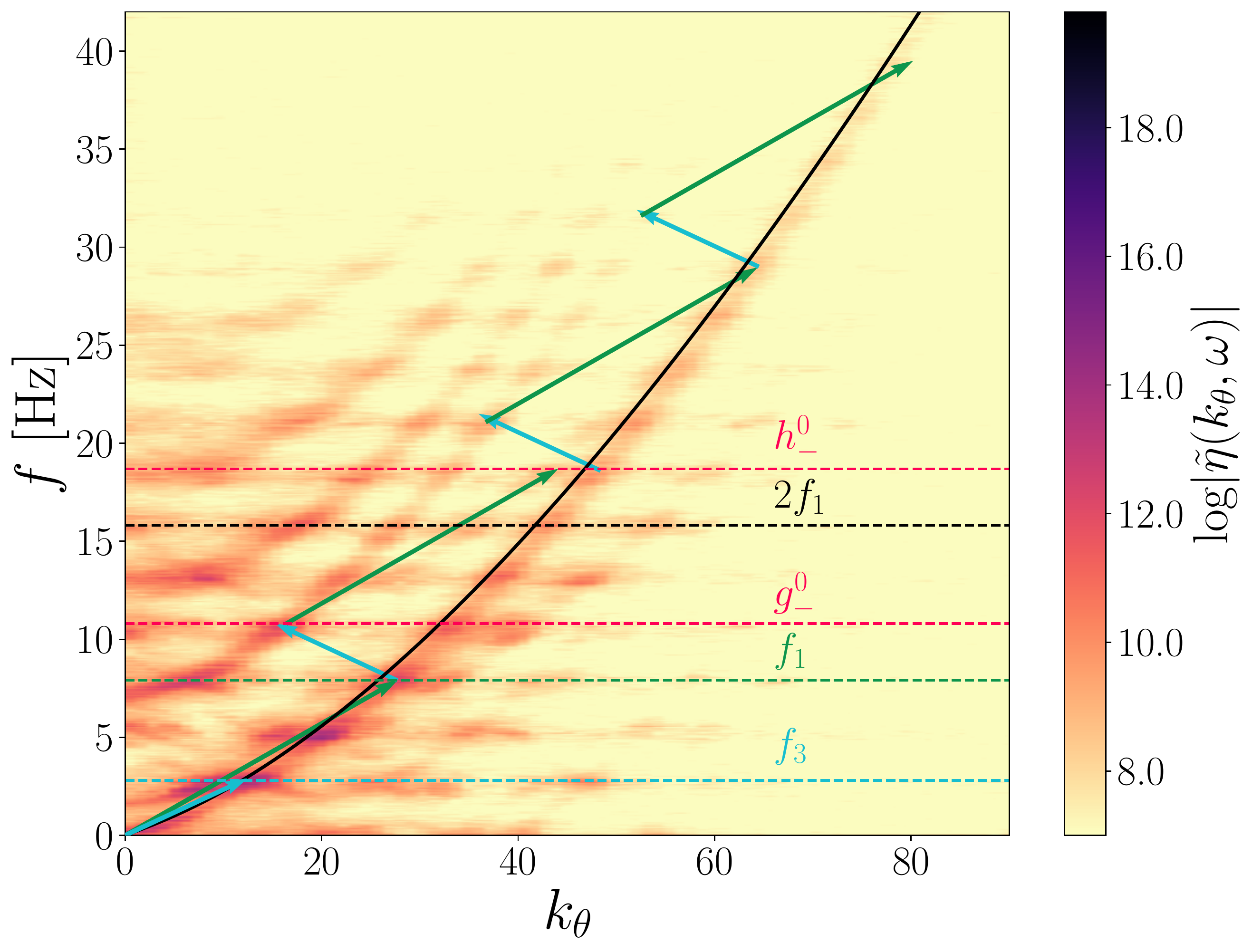}
  \caption{Fourier spectrum $\tilde{\eta}(\kth,f)$ of the torus outer border
    displacement $\ehh$. Most of the energy is concentrated along the
    gravity-capillary branch. Quasiresonant interaction is observed following
    Eq.~\eqref{eq:tertiary} and Eq.~\eqref{eq:g0k} with $\delta\kth=1.6$.
    Monochromatic forcing at $f_1=7.9$ Hz.}
\label{fig:spectrum-secondary}
\end{figure}

\subsection{Subfrequency wave generation}\label{sub:secondary}
As already noted at the bottom of Fig.~\ref{fig:spectrogram}, more than the
expected three frequencies appear in the spectrum for high enough forcings. In
order to better understand this, we apply a stronger monochromatic forcing
leading to the power spectrum in Fig.~\ref{fig:combinations}. As shown above,
the forcing frequency $f_1$, through the TRI, creates two daughters at $f_2$ and
$f_3$. Under sufficiently strong forcing, these two daughters, through
three-wave interactions create a second pair of frequencies which satisfies
\begin{align}\label{eq:couples}
  \begin{split}
  f_2-f_3&=f_-^{(1)}{\rm \,,} \\
  f_-^{(1)}+f_+^{(1)}&=f_1{\rm \,,}
  \end{split}
\end{align}
where $f_{\pm}^{(1)}$ is the first generation of secondary waves, the superscript indicating the
generation order and the subscript sign indicates the relative value. The notation $+$ indicates the
largest of the pair $f_{\pm}$ and vice-versa. The relationship between the frequencies, governed by
Eq.~\eqref{eq:couples}, is well verified experimentally in Fig.~\ref{fig:combinations}. These
grand-daughters can go on generating another generation $f^{(2)}_\pm$ in exactly the same way, which
can then be repeated again and so on, thus populating the region with a high number of discrete
peaks (see Fig.~\ref{fig:combinations}). This mechanism (analogous to that described for internal
waves~\cite{Brouzet2016}) leads to a discrete type of energy cascade, where energy is transmitted
into all the different possible daughter-wave generations. More generally, we have for the $n$th
wave generation
\begin{align}
  \begin{split}\label{eq:secondary}
  f_2-f_3&=f_-^{(n)}{\rm \,,} \\
  f_-^{(n)}+f_+^{(n)}&=f_1{\rm \,,}
  \end{split}
\end{align}
as also well observed in Fig.~\ref{fig:combinations}.

\subsection{Upper-frequency wave generation}
Let us now focus on the high-frequency part of the spectrum ($f_1<f<2f_1$) in
Fig.~\ref{fig:combinations}. The corresponding discrete set of peaks is formed
in a way similar to that of in Sec.~\ref{sub:secondary}, but involving
interaction with $f_1$. We find that these tertiary waves arise from the
interaction of the daughter waves ($f_{2,3}$) or of the secondary waves
($f^{(n)}_\pm$) with the mother wave $f_1$. We find that they satisfy the
following conditions:
\begin{align}\label{eq:tertiary}
  \begin{split}
  f_{2,3}+f_1&=g^{(0)}_{\pm} {\rm \,,}\\
  f^{(n)}_{\pm}+f_1&=g^{(n)}_{\pm}{\rm \,.}
  \end{split}
\end{align}
Note that the zeroth generation ($g^0_\pm$) is determined by the daughter waves
$f_2$ and $f_3$, whereas the $n$th generation ($n>0$) involves the secondary
waves. This relationship is verified in Fig.~\ref{fig:combinations}. Such
interaction thus provides a way to populate the high-frequency content of the
spectrum with discretely excited modes. The generation mechanism
of Eq.~\eqref{eq:tertiary} is further iterated, e.g., $g^{(0)}_\pm+f_1=h^{(0)}_{\pm}$,
$g^{(n)}_\pm+f_1=h^{(n)}_{\pm}$, as observed experimentally (not shown in Fig.~\ref{fig:combinations}).

To determine whether tertiary waves lie on the dispersion relation and
are also resonant in wavenumber we compute the experimental space-time Fourier
spectrum $\tilde{\eta}(\kth,f)$ of the torus outer border displacement as
shown in Fig.~\ref{fig:spectrum-secondary}. First, we can indeed observe that
the excited tertiary waves lie on the dispersion relation. Interestingly, for a
given tertiary wave, all branches present at that frequency are excited. But,
looking more carefully and taking $g_-^0$ as an example, we find that
 \begin{align}\label{eq:g0k}
k_\Sigma\left(g_-^0\right)&=k_{{\mathrm{gc}}}(f_1)+k_{{\mathrm{gc}}}(f_3)+\delta\kth{\rm \,,} \\
g_{-}^0&=f_1+f_3{\rm \,,}
\end{align}
 where $\delta\kth$ corresponds to the widening of the gravity-capillary dispersion branch due to
nonlinearity. $\delta\kth$ is inferred from the standard deviation of a Gaussian fit around the peak
of the Fourier spectrum at a fixed $f$. Equation~\eqref{eq:g0k} implies that a quasiresonant
interaction occurs in wavenumber.  We observed this in Fig.~\ref{fig:spectrum-secondary}, since the
frequencies have a perfect match, whereas the wavenumbers require broadening to fall on the
dispersion relation.

\subsection{Bicoherence}
Finally, we experimentally quantify the three-wave interactions (i.e.,
$\nu_1+\nu_2=\nu_3$) by computing the normalized third-order correlation in
frequency of the wave elevation called bicoherence~\cite{Kravtchenko1995}
\begin{align}
  B(\nu_1,\nu_2)=\frac{\left|\left\langle\tilde{\eta}^*(\nu_1)\tilde{\eta}^*(\nu_2)\tilde{\eta}(\nu_1+\nu_2)\right\rangle\right|}{\sqrt{\langle|\tilde{\eta}(\nu_1)\tilde{\eta}(\nu_2)|^2\rangle\langle|\tilde{\eta}(\nu_1+\nu_2)|^2\rangle}}{\rm \,,}
\end{align}
where $*$ denotes the complex conjugate. $\langle \cdot \rangle$ corresponds to an ensemble average over 101 temporal windows of the signal. The normalization is such that $B\in[0,1]$ where $0$ represents no correlation and $1$ a perfect correlation.

\begin{figure}[t!]
  \includegraphics[width=\columnwidth]{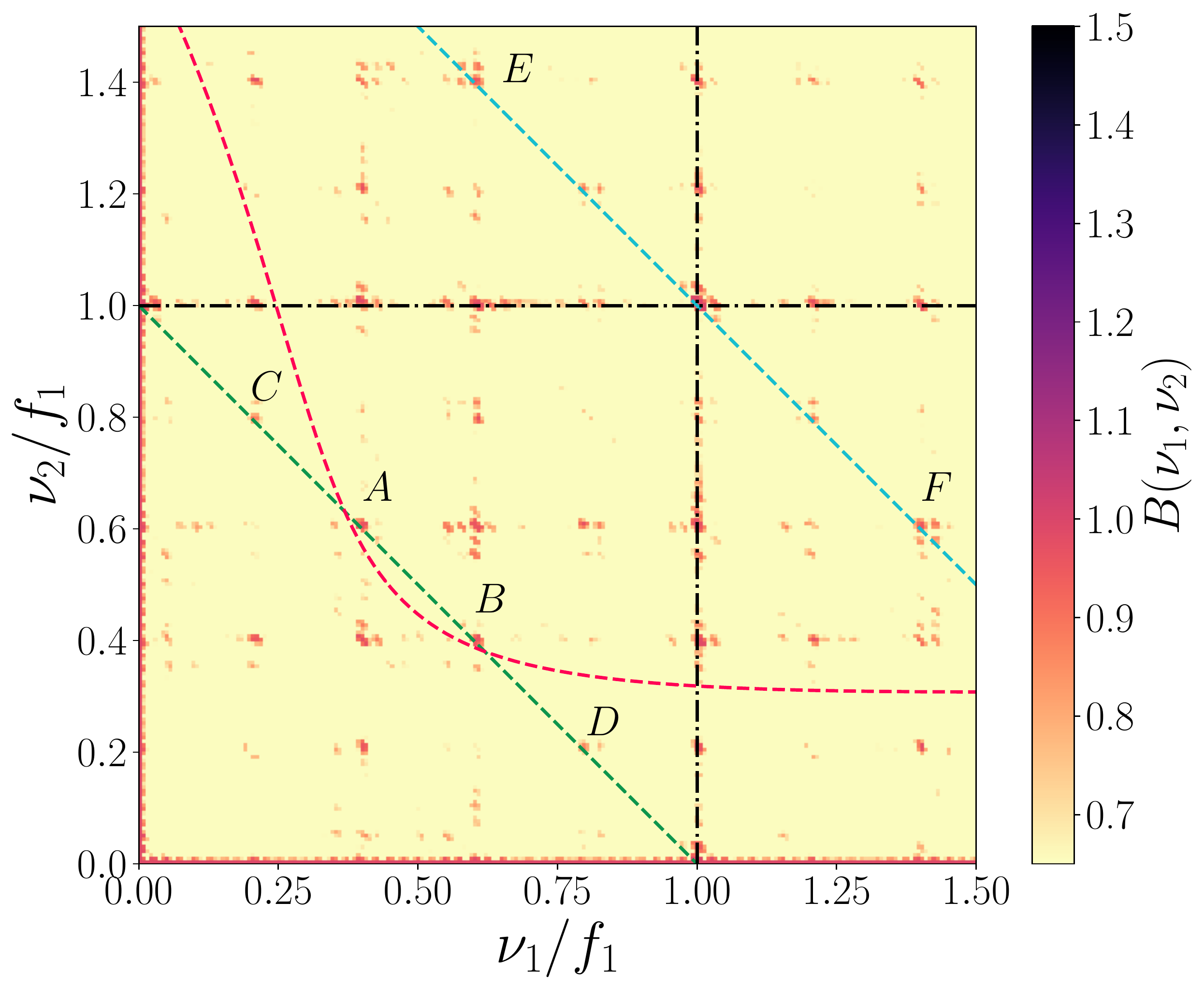}
  \caption{Bicoherence $B(\nu_1,\nu_2)$ of a wave elevation signal, recorded
    during $40$ min, at a given point. Forcing: sine wave with $f_1=7.5$ Hz. The
    mother wave is located at $(1,1)$, while all possible pairs whose sum is $1$
    are on the green oblique dashed line of slope $-1$. The two daughter waves
    are located at the intersection of this line with the resonant manifold
    (dashed red line), solutions of Eq.~\eqref{eq:resonance}, at points
    $A(f_2,f_3)$ and $B(f_3,f_2)$. Points $C$ and $D$ satisfy Eq.~\eqref{eq:couples}b whereas points $E$ and $F$ follow Eq.~\eqref{eq:tertiary}b. }
\label{fig:bicoherence}
\end{figure}

The bicoherence for a monochromatic forcing at $7.5$ Hz is depicted in
Fig.~\ref{fig:bicoherence}. We observe the primary mother wave at $(1,1)$. Note
that $B(\nu_1,\nu_2)$ is symmetric about the $\nu_2=\nu_1$ diagonal. Moreover,
the green dashed line shows all of the frequency pairs whose sum is $f_1$, but
not all of them form resonant triads.  The resonant triad $f_1=f_2+f_3$ is only
found to occur at the points $A$ and $B$, which are the intersection points
between the green dashed line and the red dashed line coming from solving the
resonance conditions of Eq.~\eqref{eq:resonance}. We can see that the daughters
are located on the intersection of the resonant manifold with the frequency-sum
line.


We can see that the plane is populated by other points, some of which are trivial (e.g., $f_1,f_1$),
as well as secondary and tertiary waves. According to Eq.~\eqref{eq:couples}b, secondary waves will be located on the green dashed line since their sum yields $f_1$. The first generation $f^{(1)}_\pm$ is found at points $C$, and by symmetry, $D$.

Tertiary waves, however, can be seen to satisfy
$g_\pm^{(n)}+f^{(n)}_\mp=2f_1$ according to Eq.~\eqref{eq:tertiary}b and using Eq.~\eqref{eq:secondary}b. This is
evidenced in Fig.~\ref{fig:bicoherence} by points E and F, lying on the cyan
dashed line which contains all points whose sum is equal to $2f_1$.

\section{Conclusion}\label{conclusion}
We have demonstrated the existence of nonlinear three-wave resonant interactions
occurring between two different branches of the hydrodynamic wave dispersion
relation, namely the gravity-capillary and sloshing modes. To the best of our
knowledge, this three-wave two-branch interaction mechanism has never been
reported experimentally in any wave system.

The system used is a torus of fluid for which the sloshing mode can easily be
excited. When subjected to a weak monochromatic forcing, a triadic resonance
instability is first observed with an exponential growth of the daughter waves
and a phase locking of the three waves. The efficiency of this interaction is
found to be maximum when the gravity-capillary phase velocity matches the group
velocity of the sloshing mode. The interaction between waves belonging to these two
branches can be considered as an analog of a forced harmonic oscillator. For
stronger forcing, additional waves are generated by a cascade of three-wave
interactions populating the high-frequency part of the wave spectrum. Since this
mechanism authorizes three-wave interactions in a 1D system far from the
gravity-capillary transition, it thus paves the way to reach a wave turbulence
regime triggered by this atypical mechanism.

In the future, we plan to explore the role of the system periodicity on the wave
interactions and on a possible wave turbulence regime, as previously shown for
solitons~\cite{Novkoski2022}. Finally, such a three-wave two-branch interaction
mechanism is probably not restricted to hydrodynamics and could be of primary
interest in other fields involving several propagation modes, such as elastic
plates~\cite{Laurent2020}, or optical waveguides~\cite{Ghatak1998}.

\begin{acknowledgments}
  We thank A. Di Palma, and Y. Le Goas for technical help. This work is supported by the French
  National Research Agency (ANR SOGOOD project No. ANR-21-CE30-0061-04), and by the Simons
  Foundation MPS No. 651463-Wave Turbulence (USA).
\end{acknowledgments}


\end{document}